\documentstyle[12pt]{article}

\begin{document}

\begin{titlepage}
\hspace{10truecm} Imperial/TP/93-94/1
\vspace{1 truecm}

\begin{center}
                {\large\bf Prima Facie Questions in Quantum Gravity
\footnote{Lecture given at the WE-Heraeus-Seminar ``The Canonical
Formalism in Classical and Quantum General Relativity'', Bad Honnef,
Germany, September 1993. Based on a lecture given at the ``Seminar in
Memory of David Bohm'', London, May 1993; and lectures given at the UK
Institute for Particle Physics, St.~Andrews, Scotland, September 1993.}
                }
\end{center}

\vspace{1 truecm}
\begin{center}
        C.J.~Isham \\[0.5cm]
        Blackett Laboratory\\
        Imperial College\\
        South Kensington\\
        London SW7 2BZ\\
        United Kingdom
\end{center}

\begin{center} October 1993\end{center}
\vspace{1truecm}

\begin{abstract}
The long history of the study of quantum gravity has thrown up a
complex web of ideas and approaches. The aim of this article is to
unravel this web a little by analysing some of the {\em prima facie\/}
questions that can be asked of almost any approach to quantum gravity
and whose answers assist in classifying the different schemes.
Particular emphasis is placed on (i) the role of background conceptual
and technical structure; (ii) the role of spacetime diffeomorphisms;
and (iii) the problem of time.
\end{abstract}
\end{titlepage}

% First a few macros
\newcommand{\be}{\begin{enumerate}}
\newcommand{\ee}{\end{enumerate}}
\newcommand{\beq}{\begin{equation}}
\newcommand{\eeq}{\end{equation}}
\newcommand{\bi}{\begin{itemize}}
\newcommand{\ei}{\end{itemize}}

\newcommand{\ie}{{\em i.e.},\ }
\newcommand{\etc}{{\em etc}}
\newcommand{\g}{g}                          % Lorentzian metric
\newcommand{\M}{M}                          % A spacetime manifold
\newcommand{\DM}{{\rm Diff}(M)}
\newcommand{\R}{{\rm I\! R}}                % The real number system

\section{Introduction}
\subsection{Preliminary Remarks}
The many ideas and suggestions that have become attached to the study
of quantum gravity form a substantial web whose subtle
interconnections often cause considerable confusion, particularly
amongst those approaching the subject for the first time. Therefore,
when seeking to assess any particular scheme (such as canonical
quantisation), it is helpful to begin by looking at the subject in the
broadest possible terms with the aim of unravelling this web a little.

    The present essay is intended to serve this need via an
examination of certain {\em prima facie\/} questions that can be used
to clarify the different structural and conceptual frameworks adopted
by the various approaches to quantum gravity. The paper starts with a
brief discussion of the four general ways whereby a quantum theory of
gravity might be constructed. This is followed by some motivation for
studying the subject in the first place and an explanation of what is
meant by a `{\em prima facie\/}' question. Next there is a short sketch
of the major current research programmes in quantum gravity: this
ensures that the subsequent discussion does not take place in a
complete technical vacuum. Then we move to three questions that are of
exceptional importance in quantum gravity: (i) the role of background
structure; (ii) the role of the spacetime diffeomorphism group; and
(iii) the `problem of time'. Finally, several of the current
approaches to quantum gravity are used to illustrate some of the
different ways in which these fundamental issues can be addressed.

    What follows is a pedagogical exposition of some basic ideas in
quantum gravity. It is {\em not\/} a full review of the field and,
for this reason, references to original work are limited in number.
Comprehensive reference lists can be found in recent genuine reviews
of quantum gravity: for example  \cite{Alv89},
\cite{Kuc92a} and \cite{Ish92,Ish93}.

\subsection{What is Quantum Gravity?}
Research in quantum gravity could perhaps be defined as any attempt
to construct a theoretical scheme in which ideas from general
relativity and quantum theory appear together in some way. A
fundamental property of any such scheme is the existence of units
with {\em dimensions\/} formed from Newton's constant $G$,
Planck's constant $\hbar$, and the ubiquitous speed of light $c$.
For example
 \bi
    \item the Planck length $L_P:=(G\hbar/c^3)^{1/2}\simeq 10^{-33}cm$;
    \item the Planck time $T_P:=L_P/c\simeq 10^{-42}s$;
    \item the Planck mass $M_P:={\hbar/cL_P}\simeq 10^{-5}gms$;
    \item the Planck energy $E_P=M_Pc^2\simeq 10^{18} GeV$;
\ei
A major question in any approach to quantum gravity is what role is
played by these fundamental units.

    This definition of quantum gravity is very broad and includes, for
example, studies of a quantum field propagating in a spacetime
manifold equipped with a fixed background Lorentzian metric. However,
in practice, references to `quantum gravity' usually include the idea
that a quantum interaction of the gravitational field with itself is
involved in some way; quantum field theory in a fixed background is
then better regarded as a way of probing certain aspects of quantum
gravity proper.

    Understood in this more limited sense, attempts to construct a
quantum theory of gravity can be divided into four broad categories
that I shall refer to as type I, type II, type III, and type IV.

\medskip\noindent
{\bf I.} The {\em quantisation\/} of general relativity.

    The idea is to start with the classical theory of general
relativity and then to apply some type of quantisation algorithm. This
is intended to be analogous to the way in which the classical theory
of an atom bound by the Coulomb potential is `quantised' by replacing
certain classical observables with self-adjoint operators on a Hilbert
space. Of course, this is essentially also the approach used in
developing important elementary-particle physics ideas like the
Salam-Weinberg electro-weak theory and the quantum chromodynamics
description of the strong nuclear force.

    Approaches to quantum gravity of this type have been studied
extensively and divide into two main categories: (i) `canonical'
schemes that start with a pre-quantum division of four-dimensional
spacetime into a three-dimensional space plus time; and (ii)
`covariant' schemes that try to apply quantum ideas in a full
spacetime context. The perturbative non-renormalisability of covariant
quantum gravity was proved in the early 1970s, and most activity in
this area ceased thereafter. On the other hand, the major advance
inaugurated by Ashtekar's work has lead to the canonical approach
becoming one of the most active branches of quantum gravity research
\cite{Ash91}.

\medskip\noindent
{\bf II.} {\em General-relativise\/} quantum theory.

    Schemes of this type are much rarer than those of type I. The
main idea is to begin with some prior idea of quantum theory and
then to force it to be compatible with general relativity. The biggest
programme of this type is due to Haag and his collaborators
\cite{FH87}.

\medskip\noindent
{\bf III.} General relativity appears as a {\em low-energy limit\/}
of a theory that is formed using conventional quantum ideas but which
does {\em not\/} involve a type-I quantisation of the classical
theory of relativity.

    The dimensional nature of the basic Planck units lends credence to
the idea of a theory that could reproduce standard general relativity in
regimes whose scales are well away from that of the Planck time,
length, energy \etc.  Superstring theory is the most successful scheme
of this type and has been the subject of intense study during the last
decade.

\medskip\noindent
{\bf IV.} Both general relativity {\em and\/} standard quantum theory
appear only in certain limiting situations in the context of a theory
that starts from radically new perspectives.

    Very little is known about potential schemes of this type or,
indeed, if it is necessary to adopt such an iconoclastic position in
order to solve the problem of quantum gravity. However, the recurring
interest in such a possibility is based on the frequently espoused
view that the basic ideas behind general relativity and quantum theory
are fundamentally incompatible and that any complete reconciliation
will necessitate a total rethinking of the central categories of
space, time and matter.

\subsection{Why Should We Study Quantum Gravity?}
Notwithstanding many decades of intense work, we are still far
from having a complete quantum theory of gravity. The problem is
compounded by the total lack of any empirical data (either
observational or experimental) that is manifestly relevant to the
problem. Under these circumstances, some motivation is necessary to
explain why we should bother with the subject at all.

\bi
\item{\em We must say something\/}.  The value of the Planck length
suggests that quantum gravity should be quite irrelevant to, for
example, atomic physics. However, the non-renormalisability of
the perturbative theory means it is impossible to actually compute
these corrections, even if physical intuition suggests they will be
minute. Furthermore, no consistent theory is known in which the
gravitational field is left completely classical. Hence we
are obliged to say {\em something\/} about quantum gravity, even if
the final results will be negligible in all normal physical domains.

\item{\em Gravitational singularities\/}. The classical theory of
general relativity is notorious for the existence of unavoidable
spacetime singularities. It has long been suggested that a quantum
theory of gravity might cure this disease by some sort of `quantum
smearing'.

\item{\em Quantum cosmology\/}. A particularly interesting singularity
is that at the beginning of a cosmological model described by, say, a
Robertson-Walker metric. Classical physics breaks down here, but one
of the aims of quantum gravity has always been to describe the
`origin' of the universe as some type of quantum event.

\item {\em The end state of the Hawking radiation process}. One of
the most striking results involving general relativity and quantum
theory is undoubtedly Hawking's famous discovery of the quantum
thermal radiation produced by a black hole. Very little is known of
the final fate of such a system, and this is often taken to be another
task for a quantum theory of gravity.

\item {\em The unification of fundamental forces}. The weak and
electromagnetic forces are neatly unified in the Salam-Weinberg
model, and there has also been a partial unification with the strong
force. It is an attractive idea that a consistent quantum theory of
gravity {\em must\/} include a unification of all the
fundamental forces.

\item {\em The possibility of a radical change in basic physics}. The
deep incompatibilities between the basic structures of general
relativity and of quantum theory have lead many people to feel that the
construction of a consistent theory of quantum gravity requires a
profound revision of the most fundamental ideas of modern physics. The
hope of securing such a paradigm shift has always been a major reason
for studying the subject.
\ei

\subsection{What Are Prima Facie Questions?}
By a `prima facie' question I mean the type of question that can be
asked of almost any approach to quantum gravity and which is concerned
with the most basic issues in the subject. The following general
classes are of particular importance:

\medskip\noindent
1. {\em General issues concerning the relation between classical and
quantum physics\/}.  The minimal requirement of any quantum theory of
gravity is that it should reproduce classical general relativity and
standard quantum theory in the appropriate physical domains. However,
it is difficult to make general categorical statements about either
(i) what is meant by the `classical limit' of a given
quantum system; or (ii) how to construct a quantum analogue of any
given classical system: in practice, the matter is usually decided on
an {\em ad hominum\/} basis. This produces significant problems in
attempts to quantise gravity using, for example, a type-I canonical
formalism.

\medskip\noindent
2. {\em Specific issues in quantum gravity\/}. Three especially
important issues of direct relevance to quantum gravity are
as follows:
\bi
\item How relevant are the {\em spacetime\/} concepts associated with
the classical theory of general relativity?  Do the Planck length and
time signify the scales at which all normal ideas of space and time
break down?

\item To what extent is it appropriate to construct a quantum theory
of gravity using the technical and conceptual apparatus of {\em
standard\/} quantum theory? For example, the traditional `Copenhagen'
interpretation of quantum theory is often asserted to be quite
inappropriate for a theory of quantum cosmology. What should take its
place?

\item Does a consistent quantum theory of gravity necessarily involve
the {\em unification\/} of the fundamental forces of physics or is it
possible to construct a theory that involves the gravitational field
alone? This questions signals one of the key differences between
superstring theory and the Ashtekar version of canonical quantum gravity:
the latter asserts that a quantisation of pure gravity {\em is \/}
possible whereas one of the main claims of superstring theory is to
provide a scheme that encompasses all the forces.
\ei

    Most readers will probably agree that questions like the above are
likely to be relevant in any approach to constructing a quantum theory
of gravity. But there is a hidden danger that should not be
underestimated. As theoretical physicists, we are inclined towards a
simple realist philosophy that sees our professional activities in
terms of using an appropriate conceptual scheme to link what is `out
there' (the world of `actual facts') to some mathematical model.
However, one of the important lessons from the philosophy of science
is that facts and theories cannot be so neatly separated: what we call
a `fact' does not exist without some theoretical schema for organising
experimental and experiential data; and, conversely, in constructing a
theory we inevitably impose some prior idea of what we mean by a fact.

    In most branches of physics no real problem arises when handling
this interconnected triad of facts, mathematical model and bridging
conceptual framework: it has simply become part of the standard
methodology of science. However, the situation in quantum gravity is
rather different since there are no known ways of directly probing the
Planck regime. This lack of hard empirical data means that research in
the subject has tended to focus on the construction of abstract
theoretical schemes that are (i) internally consistent (in a
mathematical sense), and (ii) are compatible with some preconceived
set of concepts. This rather introspective situation helps to fuel the
recurrent debate about whether the construction of a comprehensive
theory of quantum gravity requires a preliminary fundamental
reappraisal of our standard concepts of space, time and matter, or
whether it is better to try first to construct an internally
consistent mathematical model and only then to worry about what it
`means'.

    Of course, a sensible pragmatist will strive to maintain a proper
balance between these two positions, but this ever-present tension
between conceptual framework and mathematical model does lend a
peculiar flavour to much research in the field. In particular, the
wide range of views on how to approach the subject has generated a
variety of different research programmes whose practitioners not
infrequently have difficulty in understanding what members of rival
schools are trying to do. This is one reason why it is important to
uncover as many as possible of the (possibly hidden) assumptions that
lie behind each approach: one person's `deep' problem may seem
irrelevant to another simply because the starting positions are so
different. This situation also shows how important it is to try to
find some area of physics where the theory can be tested directly. A
particularly important question in this context is whether there are
genuine quantum gravity effects at scales well below the Planck
energy. Needless to say, the answer to a question like this is itself
likely to be strongly theory dependent.

\subsection{Current Research Programmes in Quantum Gravity}
At this stage it might be helpful to give a brief account of some of
the major current research programmes in quantum gravity. I find the
following scheme particularly useful, although the plethora of topics
studied could certainly be organised in many other ways.

\medskip\noindent
A. {\em Quantum Gravity Proper\/}

	The two major current programmes that attempt to construct a
full-blown theory of quantum gravity are the Ashtekar version of
canonical quantum gravity, and superstring theory.

    Canonical quantum gravity has a long history and has been used
extensively to discuss a variety of conceptual issues including the
problem of time and the possible meaning of a quantum state of the
entire universe. However, the great difficulties that arise when
trying to make proper mathematical sense of the crucial equations (in
particular, the Wheeler-DeWitt equation) that arise in this type-I
approach eventually impose insurmountable obstacles. One of the reasons
why the Ashtekar programme is so interesting is that many of these
issues can now be reopened within the context of a mathematical
framework that is much better behaved.

	The technical and conceptual framework within which
superstring theory is currently discussed is very different from that
of canonical quantum gravity and owes far more to its origins in
elementary particle physics that it does to the classical theory of
general relativity, which arises only as a low-energy limit of the
theory. As a consequence, questions concerning the status of space and
time take on a quite different form from that in the canonical
formalism. This is to be expected of any type-III approach to quantum
gravity.

    Another significant type-I scheme is the `euclidean programme'
whose basic ingredient is a functional integral over metrics with a
Riemannian signature \cite{GH93}. This view of quantum gravity cannot
be called a `full-blown' approach to quantum gravity since, as yet,
there is no known way of making mathematical sense of such integrals;
in practice, most work involves a saddle-point, semi-classical
approximation. On the other hand, the heuristic functional integral
can easily be extended to include a sum over different manifolds,
and hence the scheme is a natural one in which to discuss topology
change; something that is rather difficult in the canonical approach.

\medskip\noindent
B. {\em Quantum Cosmology}

	One of the major reasons for studying quantum gravity is to
understand the Planck era of the very early universe. Most discussions
of quantum cosmology have employed `minisuperspace' models (sometimes
in the context of the euclidean programme) in which only a finite
number of the gravitational modes are quantised. Too much weight
should not be attached to the results of such crude approximations,
especially those (the great bulk) that use the ill-defined equations
of standard canonical quantum gravity. However, models of this type
can be valuable tools for exploring the many conceptual problems that
arise in the typical quantum-cosmology situation where one aspires to
describe the quantum state of the entire universe. For example, there
has been much discussion of the problem of time and the, not
unrelated, inapplicability of the normal Copenhagen interpretation of
quantum theory. One of the currently most active ways of tackling
these issues involves the `consistent histories' approach to quantum
theory \cite{Har93a}.

\medskip\noindent
C. {\em Model Systems}

	In addition to minisuperspace techniques many other
model systems have been studied with the aim of probing specific
aspects of the full theory. For example:
\bi
\item {\em Quantum gravity theories in three or less spacetime
dimensions\/}. Low-dimensional quantum gravity has been widely
studied. In particular, the three-dimensional theory has been
solved completely using a variety of different methods and provides
valuable information on how these might be related in general.

\item {\em Quantum field theory on a spacetime with a fixed Lorentzian
geometry\/}. There has been a substantial interest in this subject
since Hawking's discovery of black-hole radiation. It throws useful
light on certain aspects of the full theory and might also have direct
astrophysical significance.

\item {\em Semi-classical quantisation\/}. There has been much
activity in recent years devoted to the WKB approximation to the
Wheeler-DeWitt equation. This approach is unlikely to reveal
much about the Planck regime proper but, nevertheless, there have
been a number of interesting results, especially suggestions that
the techniques may yield genuine results away from the scales set by
the Planck length and time.

\item {\em Regge calculus\/}. The idea of approximating spacetime
by a simplicial complex has been of interest in both classical and
quantum gravity for a long time. The quantum results to date are
rather modest but could increase in the future with the rapid increase
in the power of computing systems.
\ei

\medskip\noindent
D. {\em Spacetime Structure at the Planck Length and Time}

	A major motivation for many who elect to study quantum gravity
is the belief that something really fundamental happens to the
structure of space and time at the Planck scale. This has inspired a
number of fragmented attempts that start {\em ab initio\/} with a
theoretical framework in which standard spacetime concepts are
radically altered. Schemes of this type include twistor theory,
various approaches to a discrete models of physics, non-commutative
geometry, quantum topology or set theory, quantum causal sets, and the
like. The main difficulty is that the starting point of programmes of
this type is so far from conventional physics that it is difficult to
get back to the mundane world of Einstein field equations in a
continuum spacetime.

\section{Prima Facie Questions in Quantum Gravity}
I wish now to consider in more detail three exceptionally important
questions that can be asked of any approach to quantum gravity. These
are: (i) what background structure is assumed?; (ii) what role is
played by the spacetime diffeomorphism group?; (iii) how is the
concept of `time' to be understood?

\subsection{Background Structure}
 The phrase `background structure' can mean several things. It can
refer to a specific choice of, say, a manifold or Lorentzian metric
that is fixed once and for all and which therefore is not itself
subject to quantum effects. For example, the early attempts to
construct a quantum theory of gravity using ideas drawn from particle
physics involved writing the spacetime metric $g_{\alpha\beta}(X)$ as
a sum $\eta_{\alpha\beta}+h_{\alpha\beta}(X)$ where
$\eta_{\alpha\beta}$ is the Minkowski metric ${\rm diag}(-1,1,1,1)$.
The idea was to quantise $h_{\alpha\beta}(X)$ using standard relativistic
quantum field theory (see section \ref{SSec:QGR}). Thus, in this
type-I approach, the topology and differential structure of spacetime
are fixed as that of the vector space $\R^4$, while the background
Minkowski metric plays a key role in the quantum field theory via its
associated Poincar\'e group of isometries.

	Analysis of such background structure is a useful aid in
classifying and distinguishing the various approaches to quantum
gravity. However, `background' can be used in another way that, if
anything, is even more important but which is frequently not articulated
in such a concrete way. I mean the entire {\em conceptual\/} and {\em
structural\/} framework within whose language any particular approach
is couched. Different approaches to quantum gravity differ
significantly in the frameworks they adopt, which causes no harm---indeed
the selection of such a framework is an essential pre-requisite for
theoretical research---provided the choice is made consciously. The
problems arise when practioners from one particular school become so
accustomed to a specific structure that it becomes, for them, an
almost {\em a priori\/} set of truths, and then they find it
impossible to understand how any other position could ever be valid.
Unfortunately, a fair number of such misunderstandings have occurred
during the history of the study of quantum gravity.

	Bearing all this in mind let us begin now to examine some of
the specific issues that concern the choice of such technical or
conceptual background structure.

\medskip\noindent
A. {\em The Use of Standard Quantum Theory}

\noindent
One of the common features of superstring theory and the Ashtekar
programme is their use of standard quantum theory. True, the normal
formalism has to be adapted to handle the constraints that appear in
both approaches, but most of the familiar apparatus is present: linear
vector space, linear operators, inner products \etc. However, there
have been recurrent suggestions that a significant change in the
formalism is necessary before embarking on a full quantum gravity
programme. For example:

\be
\item General relativity may induce an essential {\em non-linearity}
into quantum theory \cite{KFL86,Pen86,Pen87}. In particular, it
may be possible to regard the infamous `reduction of the state vector'
as a genuine dynamical process induced by interactions involving the
gravitational field. The major approaches to quantum gravity would
change radically if it was necessary to start {\em ab initio\/} with a
non-linear theory rather than, say, deriving the non-linear effects as
some type of higher-order correction.

\item {How valid are the {\em continuum\/} concepts employed in
quantum theory, in particular the use of real and complex numbers? The
idea here is roughly as follows. One reply to the question ``why do we
use real numbers in quantum theory?'' is that we want the
eigenvalues of self-adjoint operators to be real numbers because they
represent the possible results of physical measurements. And why should the
results of measurements be represented by real numbers? Because all
measurements can ultimately be reduced to the positions of a pointer
in space, and space is modelled using real numbers. In other words, in
using real or complex numbers in quantum theory we are arguably making
a prior assumption about the continuum nature of space. However, it
has often been suggested that the Planck length and time signal the
scale at which standard spacetime concepts break down, and that a more
accurate picture might be a discrete structure that looks like a
familiar differentiable manifold only in some coarse-grained sense.
But we will not be able to construct such a theory if we start with a
quantum framework in which the continuum picture has been assumed {\em
a priori\/}.

	This argument is not water-tight, but it does illustrate quite
well how potentially unwarranted assumptions can enter speculative
theoretical physics and thereby undermine the enterprise.
    }

\item The Hawking radiation from a black hole is associated with a
loss of information through the event horizon and corresponds
to what an external observer would regard as a transition
from a pure state to a mixed state.  If the idea of a potential loss of
information has to be imposed on the formalism {\em ab initio\/} it
would enforce a significant change in the current quantum gravity
programmes.
\ee

    These three possible objections to standard quantum theory are all
concerned with the mathematical structure of the subject. However,
another serious problem arises with the {\em interpretation\/} of
quantum theory, especially in the context of quantum cosmology. This
particular objection has been taken very seriously in recent years and
by now there is a fairly widespread agreement that the familiar
`Copenhagen' view is not appropriate. In particular, there is a strong
desire to find an alternative interpretation whose fundamental
ingredients do not include the notion of a measurement by an external
observer.

    The consistent-histories approach may provide such a scheme
although, even here, there is a problem in so far as most discussions
of this subject presuppose the existence of a fixed spacetime. In
conventional quantum theory there is certainly a strong case for
arguing that this is necessary, both for the mathematical foundations
and for the conceptual interpretation of the theory. This raises what
turns out to be one of the most interesting {\em prima facie\/}
questions in quantum gravity: how much of the standard spacetime
structure must be imposed as part of the fixed background?

\medskip\noindent
B. {\em How Much Spacetime Structure Must Be Fixed?}

\noindent
The mathematical model of spacetime used in classical general
relativity is a differentiable manifold equipped with a
Lorentzian metric. Some of the more important pieces of substructure
underlying this picture are illustrated in Figure 1.
\begin{figure}[t]
\begin{center}
    \fbox{\large\rule[-1ex]{0ex}{3.7ex}\ Set $\M$ of spacetime
                    points/events\ }
\end{center}

\begin{center}$\downarrow$\end{center}

\begin{center}
    \fbox{\large\rule[-1ex]{0ex}{3.7ex}\ Topological structure\ }
\end{center}

        \begin{center}$\downarrow$\end{center}

\begin{center}
    \fbox{\large\rule[-1ex]{0ex}{3.7ex}\ Manifold structure \ }
\end{center}

        \begin{center}$\downarrow$\end{center}

\begin{center}
    \fbox{\large\rule[-1ex]{0ex}{3.7ex}\ Causal structure\ }
\end{center}

        \begin{center}$\downarrow$\end{center}

\begin{center}
    \fbox{\large\rule[-1ex]{0ex}{3.7ex}\ Lorentzian structure $g$\ }
\end{center}

        \begin{center}{Figure 1. The spacetime structure of classical
                            general relativity}
\end{center}
\end{figure}

    The bottom level is a set $\M$ whose elements are to be identified
with spacetime `points' or `events'. This set is formless with its
only general mathematical property being the cardinal number. In
particular, there are no relations between the elements of $\M$ and no
special way of labelling any such element.

    The next step is to impose a topology on $\M$ so that each point
acquires a family of neighbourhoods. It now becomes possible to talk
about relationships between points, albeit in a rather non-physical
way. This defect is overcome by adding the key ingredient of all
standard views of spacetime: the topology of $\M$ must be compatible
with that of a differentiable manifold. A point can then be labelled
uniquely in $\M$ (at least, locally) by giving the values of four real
numbers. Such a coordinate system also provides a more specific way of
describing relationships between points of $\M$, albeit not
intrinsically in so far as these depend on which coordinate
systems are chosen to cover $\M$.

    In the final step a Lorentzian metric $\g$ is placed on $\M$,
thereby introducing the ideas of the lengths of a path joining two
spacetime points, parallel transport with respect to a Riemannian
connection, causal relations between pairs of points \etc. There are
also a variety of possible intermediate steps between the manifold and
Lorentzian pictures; for example, as signified in Figure~1, the idea
of a causal structure is more primitive than that of a Lorentzian
metric.

    The key question is how much of this classical structure is to
be held fixed in the quantum theory. In the context of the Copenhagen
interpretation the answer is arguably ``all of it''. I suspect that
Bohr would have identified the spacetime Lorentzian structure as an
intrinsic part of that classical world which he felt was such an
essential epistemological prerequisite for the discussion of quantum
objects. It seems probable therefore that he would not have approved
at all of the subject of quantum gravity!

    However, since we wish to assert that some sort of quantum
spacetime structure {\em is\/} meaningful, the key question for any
particular approach to quantum gravity is how {\em much\/} of the
hierarchy illustrated in Figure~1 must be kept fixed. For example, in
most type-I approaches to `quantising' the gravitational field, the
set of spacetime points, topology and differential structure are all
fixed, and only the Lorentzian metric $\g$ is subject to quantum
fluctuations. But one can envisage more interesting type-I
possibilities in which, for example, the set $\M$ and its topology are
fixed, but quantum fluctuations are permitted over all those manifold
structures that are compatible with this particular topology. Moving
back a step, one can envisage the exotic idea of fixing only the
set-theoretic structure of $\M$ and allowing quantum fluctuations over
topologies that can be placed on this set, including perhaps many that
are not compatible with a manifold structure at all. Finally, one
might even imagine `quantising' the point set $\M$ itself: presumably by
allowing quantum fluctuations in its cardinal number.

    The ideas sketched above are all examples of what might
be called `horizontal' quantisation in which the quantum fluctuations
take place only within the category of objects specified by the
classical theory. However, it is also possible to contemplate
`vertical' quantisation in which fluctuations take place in a wider
category. A simple example, and one that arises in several type-I schemes,
is to permit quantum fluctuations in the metric to include fields
that are degenerate, or have signature other than $(-1,1,1,1)$. A more
exotic possibility would be to allow fluctuations in manifold structure
that include non-commutative manifolds.

    The ideas of `horizontal' and `vertical' quantisation arise most
naturally in the context of type-I approaches in which a given
classical system is `quantised' in some way. However, the question of
how much of the classical structure of spacetime remains at different
levels of the full theory can be asked meaningfully in all the four
general approaches to quantum gravity. This is related to the question
of the role of the Planck length in such theories. A common
expectation is that the standard picture of space and time is
applicable only at scales well above the Planck regime, and that the
Planck length, time, energy \etc\ signal the point at which {\em phase
transitions\/} take place.

    The notion of different phases is attractive but it also suggests
that a complete theory of quantum gravity should assume no prior
spacetime structure at all.  Of course, this does not forbid the
construction of partial theories that describe the theory in a
particular phase; indeed, this may be a necessary first stage in
the construction of the full structure.

\medskip\noindent
C. {\em Background Causal Structure}

    A piece of potential background of particular importance
is causal structure. For example, consider the problem of constructing
the quantum theory of a scalar field $\phi$ propagating on a spacetime
manifold $\M$ equipped with a fixed Lorentzian metric $\g$. In such a
theory, a key role is played by the microcausal condition
\beq
    [\,\hat\phi(X),\hat\phi(Y)\,]=0             \label{micro_causal}
\eeq
for all spacetime points $X$ and $Y$ that are spacelike separated
with respect to $\g$. More rigorously, if ${\cal A}(O)$
denotes the $C^*$-algebra of local observables associated with the
spacetime region $O\subset\M$, then we require $[{\cal A}(O_1),{\cal
A}(O_2)]=0$ for any disjoint regions $O_1$ and $O_2$ that are
spacelike separated.

    Now consider what happens in a type-I, `covariant' approach to
quantum gravity. If the Lorentzian metric $\g$ becomes quantised
then the light cone associated with any spacetime point is no
longer fixed and it is not meaningful to impose a microcausal relation
like (\ref{micro_causal}): any pair of spacetime points are
`potentially' null or time-like separated, and hence spacetime quantum
fields can never commute.

    This collapse of one of the bedrocks of conventional quantum field
theory is probably the single greatest reason why spacetime approaches
to quantum gravity have not got as far as might have been hoped. In
the original, particle-physics based schemes the problem was
circumscribed by introducing a background Minkowskian metric $\eta$
and then quantising the graviton using the microcausal structure
associated with $\eta$. Such a background is also necessary for the
idea of `short-distance' behaviour (which plays a key role in
discussing renormalisability) to have any meaning. However, the use of
a fixed causal structure is an anathema to most general relativists
and therefore, even if this approach to quantum gravity had worked
(which it did not), there would still have been a strong compunction
to reconstruct the theory in a way that does not employ any such
background.

    This very non-trivial problem is one of the reasons why the
canonical approach to quantum gravity has been so popular. However,
a causal problem arises here too. For example, in the
Wheeler-DeWitt approach, the configuration variable of the system is
the Riemannian metric $q_{ab}(x)$ on a three-manifold $\Sigma$, and the
canonical commutation relations invariably include the set
\beq
    [\,\hat q_{ab}(x),\hat q_{cd}(x')\,]=0
\eeq
for all points $x$ and $x'$ in $\Sigma$. In normal canonical quantum
field theory such a relation arises because $\Sigma$ is a space-like
subset of spacetime, and hence the fields at $x$ and $x'$ should be
simultaneously measurable.  But how can such a relation be justified
in a theory that has no fixed causal structure?  This problem is
rarely mentioned but it means that, in this respect, the canonical
approach to quantum gravity is no better than the covariant one. It
is another aspect of the `problem of time' to which my second lecture
is devoted

\subsection{The Role of the Spacetime Diffeomorphism Group $\DM$}
The group $\DM$ of spacetime diffeomorphisms plays a key role in the
classical theory of general relativity and so the question of its status
in quantum gravity is of considerable {\em prima facie\/} interest. We
shall restrict our attention to diffeomorphisms with compact support,
by which I mean those that are equal to the unit map outside some
closed and bounded region of $\M$. Thus, for example, a
Poincar\'e-group transformation of Minkowski spacetime is not deemed
to belong to $\DM$. This restriction is imposed because the role of
transformations with a non-trivial action in the asymptotic regions of
$M$ is quite different from those that act trivially.

    The role of $\DM$ in quantum gravity depends strongly on the
approach taken to the subject. For example, in a type-III or type-IV
scheme  the structure of classical relativity is expected to appear
only in a low-energy limit and so there is no strong reason to suppose
that $\DM$ will play any fundamental role in the quantum theory. A
type-II scheme is quite different since the group of spacetime
diffeomorphisms is likely to be a key ingredient in forcing a quantum
theory to comply with the demands of general relativity. On the other
hand, the situation for type-I approaches is less clear. Any scheme
based on a prior canonical decomposition into space plus time is bound
to obscure the role of spacetime diffeomorphisms, and even in
the covariant approaches quantisation may affect enough of the
classical theory to detract from the significance of such
transformations.

    Some insight can be gained by looking at certain aspects of
spacetime diffeomorphisms in classical general relativity. It is
helpful here to distinguish between the pseudo-group of local
coordinate transformations and the genuine group $\DM$ of global
diffeomorphisms of $\M$. Compatibility with the former can be taken to
imply that the theory should be written using tensorial objects on
$\M$. On the other hand, as Einstein often emphasised, $\DM$
appears as an active group of transformations of $\M$, and invariance
under this group implies that the points in $\M$ have no direct
physical significance. Of course, this is also true in special
relativity but it is mitigated there by the existence of inertial
reference frames that can be transformed into each other by the
Poincar\'e group of isometries of the Minkowski metric.

    Put somewhat differently, the action of $\DM$ on $\M$ induces an
action on the space ${\cal F}$ of spacetime fields, and the only thing
that has immediate physical meaning is the quotient space ${\cal
F}/\DM$ of orbits, \ie two field configurations are regarded as
physically equivalent if they are connected by a $\DM$ transformation.
Technically, this is analogous to the situation in electromagnetism
whereby a vector potential $A_\mu$ is equivalent to
$A_\mu+\partial_\mu f$ for all functions $f$. However, there is an
important difference between the electromagnetism and
general-relativity. Electromagnetic gauge transformations occur at a
fixed spacetime point $X$, and the physical configurations can be
identified with the values of the electromagnetic field
$F_{\mu\nu}(X)$, which depends {\em locally\/} on points of $\M$. On
the other hand, $\DM$ maps one spacetime point into another, and
therefore the obvious way of constructing a $\DM$-invariant object is
to take some scalar function of spacetime fields and integrate it over
the whole of $\M$, which gives something that is very {\em
non\/}-local. The idea that `physical observables' are naturally
non-local is an important ingredient in many approaches to quantum
gravity.

\subsection{The Problem of Time}
One of the major issues in quantum gravity is the so-called `problem
of time'. This arises from the very different roles played by the
concept of time in quantum theory and in general relativity. Let us
start by considering standard quantum theory.
\be
    \item Time is not a physical observable in the normal sense since
it is not represented by an operator. Rather, it is treated as a {\em
background\/} parameter which, as in classical physics, is used to
mark the evolution of the system; in this sense it can be regarded as
part of Bohr's background classical structure. In particular, it
provides the parameter $t$ in the time-dependent Schr\"odinger
equation
\beq
        i\hbar{d\psi_t\over dt}= \hat H\psi_t.   \label{SE}
\eeq
This is why the meaning assigned to the time-energy uncertainty
relation $\delta t\,\delta E\ge{1\over 2}\hbar$ is quite different
from that associated with, for example, the position and the momentum
of a particle.

    \item The idea of events happening at a single time plays a
crucial role in the technical and conceptual foundations of quantum
theory:
    \bi
        \item The notion of a {\em measurement\/} made at a
particular time is a fundamental ingredient in the conventional
Copenhagen interpretation. In particular, an {\em observable\/} is
something whose value can be measured at a fixed time.

        \item One of the central requirements of the {\em scalar
product\/} on the Hilbert space of states is that it is conserved
under the time evolution (\ref{SE}). This is closely connected to the
unitarity requirement that probabilities always sum to one.

        \item More generally, a key ingredient in the construction of
the Hilbert space for a quantum system is the selection of a
complete set of observables that are required to {\em commute\/} at
a fixed value of time.
    \ei

    \item These ideas can be extended to systems that are compatible
 with special relativity: the unique time system of Newtonian physics
is simply replaced with the set of relativistic inertial reference
frames.  The quantum theory can be made independent of a choice of
frame if it carries a unitary representation of the Poincar\'e group.
In the case of a relativistic quantum field theory, this is closely
related to the microcausality requirement, which---as emphasised
earlier---becomes meaningless if the light cone is itself the subject
of quantum fluctuations.
\ee
    The background Newtonian time appears explicitly in the
time-dependent Schr\"odinger equation (\ref{SE}), but it is pertinent
to note that such a time is truly an abstraction in the sense that no
{\em physical\/} clock can provide a precise measure of it
\cite{UW89}: there is always a small probability that a real clock
will sometimes run backwards with respect to Newtonian time.

    When we come to a $\DM$-invariant theory like classical general
relativity the role of time is very different. If $\M$ is equipped with
a Lorentzian metric $g$, and if its topology is appropriate, it
can be foliated in many ways as a one-parameter family of space-like
surfaces, and each such parameter might be regarded as a possible
definition of time. However several problems arise with this
way of looking at things:
\bi
\item There are many such foliations, and there is no way of selecting
a particular one, or special family of such, that is `natural' within
the context of the theory alone.

\item Such a definition of time is rather non-physical since it
provides no hint as to how it might be measured or registered.

\item The possibility of defining time in this way is closely linked
to a fixed choice of the metric $g$. It becomes untenable if $g$ is
subject to some type of quantum fluctuation.
\ei

    The last problem is crucial in any type-I approach to quantum
gravity and raises a number of important questions. In particular:
\bi
\item How {\em is\/} the notion of time to be incorporated in a
quantum theory of gravity?

\item Does it play a fundamental role in the construction of the
theory or is it a `phenomenological' concept that applies, for example,
only in some coarse-grained, semi-classical sense?

\item In the latter case, how reliable is the use at a basic level of
techniques drawn from standard quantum theory?
\ei

    The three main ways that have been suggested for solving the problem
of time are as follows.
\be
\item Fix some {\em background\/} causal structure and use that to
determine temporal concepts in the quantum theory. Such a background
might arise from two possible sources.
    \bi
    \item It might come from a {\em contingent\/} feature of the
actual universe; for example, the $3^0$K thermal radiation. However,
structure of this type is approximate and therefore works only if fine
details are ignored. Also, there is a general matter of principle: do
we expect a quantum theory of gravity to work for `all possible'
universes (whatever that might mean), or only for the actual one in
which we happen to live?

    \item  An asymptotic causal structure could be associated with a
spacetime manifold that is spatially non-compact and asymptotically
flat. However, this would not help in the typical cosmological
situation, and it is by no means obvious that `time' defined in this
way can be measured in any physically meaningful way.
    \ei

\item {Attempt to locate events both spatially and temporally with
specific functionals of the gravitational and other fields. This
important idea is based on the observation that, for example, if
$\phi$ is a scalar field then, as emphasised earlier, the value
$\phi(X)$ of $\phi$ at a particular $X\in M$ has no physical
meaning. On the other hand, the value of $\phi$ where something `is'
{\em does\/} have a physical meaning in the sense that `$\phi({\rm
thing})$' is $\DM$ invariant.

    The hope is that an `internal time' of this type can be introduced
in such a way that the normal dynamical equations of the classical
theory are reproduced {\em precisely\/}. Then one would try to apply a
similar technique to the quantum case. Ideas of this type have
played a major role in the development of canonical quantum gravity.
    }

\item The third approach starts by constructing some sort of quantum
theory but with no reference to time at all. Physical time is then
introduced as a reading on `real clocks' but it is accepted that such
a scheme will never {\em exactly\/} reproduce the standard notion of
time and that all physical clocks will at best work in some
semi-classical limit. Approaches of this type are truly `timeless' and
raise the key issue of whether a meaningful quantum theory can indeed
be created in a way that contains no fundamental reference to time.
That this is not a trivial matter is exemplified by the remarks made
earlier about the crucial role of time in conventional quantum theory.
\ee

\section{Approaches to Quantum Gravity}
As explained earlier, there are four general ways of trying to
construct a quantum theory of gravity: I.,\ quantize general
relativity; II.,\ general-relativise quantum theory; III.,\ schemes
constructed using standard quantum theory in which general relativity
emerges only in some low-energy limit; and IV.,\ schemes in which both
general relativity and quantum theory emerge in some appropriate
domain in the context of a theory that contains radical new ideas. Our
task now is to see how the {\em prima facie\/} issues discussed in the
previous section are addressed in some of these schemes.

\subsection{Quantize General Relativity}
\label{SSec:QGR}
\noindent
A. {\em The Particle-Physics Approach}

\noindent
The early particle-physics based approaches to quantum gravity
illustrate quite well a number of the issues discussed above. The
starting point is to fix the background topology and differential
structure of spacetime $\M$ to be that of Minkowski space, and then to
write the Lorentzian metric $g$ on $\M$ as
\beq
        g_{\alpha\beta}(X)=\eta_{\alpha\beta}+h_{\alpha\beta}(X)
                                        \label{g=eta+h}
\eeq
where $h$ measures the departure of $g$ from flat spacetime $\eta$.

    The background metric $\eta$ provides a fixed causal structure
with the usual family of Lorentzian inertial frames. Thus, at this
level, there is no problem of time. The causal structure also allows
a notion of microcausality, thereby permitting a conventional type
of relativistic quantum field theory to be applied to the field
$h_{\alpha\beta}$. In particular, the quanta of this field (defined
as usual using representations of the Poincar\'e group of isometries
of $\eta$) are massless spin-2 particles. A typical task would then
be to compute perturbative scattering-matrix elements for these
gravitons, both with each other and with the quanta of various
matter fields. Note that there is no immediate problem of
interpretation: the existence of a background spacetime manifold and
causal structure fits in well with the standard Copenhagen
view of quantum theory.

    The action of $\DM$ is usually studied infinitesimally and is
reflected in the quantum theory via a set of Ward identities that
must be satisfied by the $n$-point functions of the theory. Thus the
role of spacetime diffeomorphisms is also relatively
straightforward.

    It is clear that many of the {\em prima facie\/} issues discussed
earlier are resolved in an approach of this type by virtue of its
heavy use of background structure. However, many classical relativists
object violently to an expansion like (\ref{g=eta+h}), not least
because the background   \footnote{The scheme can be extended to use
an arbitrary background metric, but this does not change the force of
the objection.}  causal structure cannot generally be identified with
the physical one. Also, one is restricted to a specific background
topology, and so a scheme of this type is not well adapted for
addressing many of the most interesting questions in quantum gravity:
black-hole phenomena, quantum cosmology, phase changes \etc.
Nevertheless, if the scheme above had worked it
would have been a major result and would undoubtedly have triggered a
substantial effort to construct a covariant type-I theory in a
non-perturbative way; a good analogue is the great increase in
studies of lattice gauge theory that followed the proof by t'Hooft
that Yang-Mills theory is perturbatively renormalisable.

    Of course, this did not happen in the gravitational case because
the ultraviolet divergences are sufficiently violent to render the
theory perturbatively non-renormalisable. One reaction has been to
regard this pathology as a result of using the expansion
(\ref{g=eta+h}); an expansion that is, anyway, unpleasant when viewed
from the canons of the classical theory. Several attempts have been
made to construct a non-perturbative, covariant scheme, but none is
particularly successful and it was only when Ashtekar made his
important discoveries in the context of the {\em canonical\/} theory
that the idea of non-perturbative quantisation really began to bear
fruit.

    The majority of particle physicists followed a different line and
tried to enlarge the classical theory of general relativity with
carefully chosen matter fields with the hope that the ultraviolet
divergences would cancel, leaving a theory that is perturbatively
well-behaved. The cancellation of a divergence associated with a loop
of bosonic particles (like the graviton) can be achieved only by the
introduction of {\em fermions\/}, and hence supergravity was born.
However, supersymmetry requires very special types of matter, which
supports the idea that a successful theory of quantum gravity must lead
to a unified theory; \ie the extra fields needed to cancel the
graviton infinities might be precisely those associated with some
grand unified scheme of the fundamental forces.

    Early expectations were high following successful
low-order results but it is now generally accepted that if
higher-loop calculations could be performed (they are very complex)
intractable divergences would appear once more. However, this line of
thinking is far from dead and the torch is currently carried by
perturbative superstring theory.

    Superstring theory has the great advantage over the simple
covariant approaches that the individual terms in the appropriate
perturbation expansion {\em can\/} be finite and, furthermore, the
particle content of theories of this type could well be such as to
relate the fundamental forces in a unified way.

    The low-energy limit of these theories is a form of supergravity
but, nevertheless, standard spacetime ideas do not play a very
significant role. This is reflected by the graviton being only one
of an infinite number of particles in the theory; similarly,
the spacetime diffeomorphism group appears only as part of a much
bigger structure. This down-playing of classical general relativity is
typical of a type-III approach.

    Notwithstanding the successes of superstring theory, some of the
earlier objections to perturbative schemes still hold and, in
addition, the superstring perturbation series is highly divergent.
Hence much current attention is being devoted to  the challenge of
constructing a non-perturbative version of the theory. Most of the
suggestions made so far work within the context of standard quantum
theory and, in this sense, they are still of type III. However, the
possibility also arises of finding a genuine type-IV structure whose
low-energy limits would include standard quantum theory as well as
supergravity.

\medskip\noindent
B. {\em The Canonical Approach}

\noindent
The response to our {\em prima facie\/} questions given by the
canonical approach to quantum gravity differs significantly from
that of the particle-physics based schemes. Since my second lecture
is devoted to canonical quantisation, I will merely sketch here some
of the most important features.

\smallskip\noindent
1. {\em Use of standard quantum theory\/}. The basic technical ideas
of standard quantum theory are employed, albeit adapted to handle the
non-linear constraints satisfied by the canonical variables. On the
other hand the traditional, Copenhagen type of interpretation of
the theory is certainly {\em not\/} applicable in quantum
cosmology, which is one of the most important potential uses of
canonical quantum gravity.

\smallskip\noindent
2. {\em Background manifold stucture\/}. The canonical theory of
classical relativity assumes {\em ab initio\/} that the spacetime
manifold $\M$ is diffeomorphic to $\Sigma\times\R$ where $\Sigma$ is
some three-manifold. This three-manifold becomes part of the fixed
background in the quantum theory and so, for example, there is no
immediate possibility of discussing quantum topology.

\smallskip\noindent
3. {\em Background metric structure\/}. One of the main aspirations of
the canonical approach to quantum gravity has always been to build a
formalism with no background spatial, or spacetime, metric. This is
particularly important in the context of quantum cosmology.

\smallskip\noindent
4. {\em The spacetime diffeomorphism group\/}. In the canonical form
of general relativity the spacetime diffeomorphism group $\DM$ is
replaced by a more complex entity (the `Dirac algebra') which contains
${\rm Diff}(\Sigma)$ as a subgroup but which is not itself a genuine group.
Invariance under ${\rm Diff\/}(\Sigma)$ means that the functionals of the
canonical variables that correspond to physical variables are
non-local with respect to $\Sigma$. The role of the full Dirac algebra
is more subtle and varies according to which canonical scheme is
followed.

\smallskip\noindent
5. {\em The problem of time\/}. In the absence of any background
metric, this becomes a major issue. It is closely connected with the
role of the Dirac algebra and the general question of what is meant
by an `observable'. None of the several suggested ways for handling
this problem work fully and it seems plausible that the standard concept of
time can be recovered only in some semi-classical sense.

\smallskip\noindent
6. {\em The unification of the fundamental forces\/}. Perturbation
theory in simple canonical quantum gravity is as badly defined
mathematically as is its particle-physics based cousin. However,
developments in the Ashtekar programme imply that it may be
possible to construct a non-perturbative theory that is finite and
that involves just the gravitational field alone. In this sense,
canonical quantum gravity does {\em not\/} suggest that a
unification of the forces is a necessary ingredient of a technically
successful theory.

\subsection{General-Relativise Quantum Theory\/}
I cannot say much about the idea `general-relativising' quantum
theory (\ie type-II schemes) in relation to our {\em prima facie\/}
issues because little research has been done in this area. A key
role would probably be played by the spacetime diffeomorphism group
$\DM$: indeed, a type-II scheme might be {\em defined\/} as any
attempt to force standard quantum theory to be compatible with
$\DM$; of course the structure of $\M$ itself would then necessarily
be part of the fixed background.

    Important questions that arise in an approach of this type
include:
\bi
\item What form of quantum theory should be used? In particular, does
it require a prior notion of `time'?

\item What is the role of the field equations of classical general
relativity? Do they also need to be imposed as part of the structural
background, or is the fundamental input the spacetime
diffeomorphism group alone?

\item Is there a `canonical' version in which standard quantum theory
is forced to be compatible with the Dirac algebra rather than with
$\DM$?

\item There is a long-standing and extensive research programme to
construct quantum field theories (usually linear) in a  background
spacetime manifold equipped with a fixed Lorentzian metric. Does this
work throw any light on the idea of general-relativising quantum theory?
\ei

\subsection{The Use of Radical New Concepts}
In the classification we have been using, a type-IV scheme is any
approach to quantum gravity that starts with a view of quantum theory
and spacetime physics that is radically different from that of
conventional theories, and with the expectation that these
standard ideas will emerge only in some limited domain. Almost by
definition, schemes of this type dispense with much of the background
structure of other approaches to quantum gravity, including, possibly,
standard quantum theory as well as many normal spacetime concepts.
Unfortunately, such schemes tend to be individualistic in form, and
their manner of dealing with our {\em prima facie\/} issues has to be
treated on a strictly {\em ad hominum\/} basis.

\section{Conclusions}
The two major current approaches to quantum gravity proper---the
Ashtekar programme, and superstring theory---differ so much in their
starting positions and lines of development that it is hard to say
much in conclusion other than that the problem of quantum gravity is
still wide open. In particular, and {\em pace\/} the discussion above,
the jury is still out on the all-important question of whether a
consistent theory of quantum gravity can be achieved within the
framework of our existing understanding of physics, or whether some
radical change is needed before any real headway can be made.

    The problem of time is crucial in this respect, and its resolution
is still very unclear. However. one reading of the current situation
is that normal notions of time and space are applicable only at scales
well above the Planck regime. If true, such a position throws great
doubt on the use of {\em any\/} standard quantum ideas as a basic
ingredient in the theory; indeed, a more plausible scenario is that
standard quantum theory becomes applicable at precisely the same point
in the formalism as does the normal notion of time. Such a situation
is exciting for those who, like myself, enjoy indulging in speculative
metaphysics/theoretical-physics, but it is also most frustrating in
the absence of any clear empirical data that could point us in the
right direction. The problem of quantum gravity continues to be a
challenge for the next century!

%%%%%%%%%%%%%%%%%%%%%%%%%%%%%%%%%%%%%%%%%%%%%%%%%%%%%%%%%%%%%%%%%%%%%%%%

\end{document}